\documentclass[letterpaper,12pt]{JHEP3}
\usepackage{amsmath,amssymb}
\usepackage{latexsym}
\usepackage{epsfig}
\usepackage{cite}
\usepackage[english]{babel}

\newcommand{\eq}{\begin{equation}}
\newcommand{\feq}{\end{equation}}
\newcommand{\eqn}{\begin{eqnarray}}
\newcommand{\feqn}{\end{eqnarray}}
\newcommand{\arr}{\begin{eqnarray*}}
\newcommand{\farr}{\end{eqnarray*}}

\font\mybb=msbm10 at 12pt
\def\bb#1{\hbox{\mybb#1}}

\def\bR {\bb{R}}

\def\bC {\bb{C}}

\title{Nut-charged black holes in matter-coupled  ${\cal N}=2$, $D=4$ gauged supergravity}

\author{Marta Colleoni$^a$ and Dietmar Klemm$^{ab}$ \\
$^a$ Dipartimento di Fisica, Universit\`a di Milano, \\
\hspace*{0.15cm} Via Celoria 16, I-20133 Milano. \\
$^b$ INFN, Sezione di Milano, Via Celoria 16, I-20133 Milano. \\
}
\preprint{IFUM-991-FT}

\abstract{Using the results of arXiv:0804.0009, where all
timelike supersymmetric backgrounds of ${\cal N}=2$, $D=4$ matter-coupled supergravity
with Fayet-Iliopoulos gauging were classified, we construct
genuine nut-charged BPS black holes in AdS$_4$ with nonconstant moduli. The calculations are
exemplified for the $\text{SU}(1,1)/\text{U}(1)$ model
with prepotential $F=-iX^0X^1$. The resulting supersymmetric black holes have a hyperbolic
horizon and carry two electric,
two magnetic and one nut charge, which are however not all independent, but are given in terms of
three free parameters. We find that turning on a nut charge lifts the flat directions in the effective black
hole potential, such that the horizon values of the scalars are completely fixed by the charges.
We also oxidize the solutions to eleven dimensions, and find that they generalize the geometry found in
hep-th/0105250 corresponding to membranes wrapping holomorphic curves in a Calabi-Yau
five-fold. Finally, a class of nut-charged Nernst branes is constructed as well, but these have curvature
singularities at the horizon.
}


\begin{document}

\section{Introduction}
\label{intro}

Black holes in anti-de~Sitter (AdS) spaces provide an important testground
to address fundamental questions of quantum gravity like holography. These
ideas originally emerged from string theory, but became
then interesting in their own right, for instance in recent applications to condensed
matter physics (cf.~\cite{Hartnoll:2009sz} for a review), where black holes
play again an essential role, since they provide the dual description of certain condensed
matter systems at finite temperature. In particular, models of the type that we shall consider here,
that contain Einstein gravity coupled to $\text{U}(1)$ gauge fields and neutral scalars\footnote{The
necessity of a bulk $\text{U}(1)$ gauge field arises, because a basic ingredient of realistic condensed
matter systems is the presence of a finite density of charge carriers. A further step in modeling strongly
coupled holographic systems is to include the leading relevant (scalar) operator in the dynamics.
This is generically uncharged, and is dual to a neutral scalar field in the bulk.}
have been instrumental to study transitions from Fermi-liquid to
non-Fermi-liquid behaviour, cf.~\cite{Charmousis:2010zz,Iizuka:2011hg} and references therein.

On the other hand, among the extremal black holes (which have zero Hawking
temperature), those preserving a sufficient amount of supersymmetry are of
particular interest, as this allows (owing to non-renormalization theorems) to
extrapolate an entropy computation at weak string coupling (when the system
is generically described by a bound state of strings and branes) to the
strong-coupling regime, where a description in terms of a black hole is
valid \cite{Strominger:1996sh}. However, this picture, which has been essential
for our current understanding of black hole microstates, might be modified in
gauged supergravity (arising from flux compactifications) due to the presence of
a potential for the moduli, generated by the fluxes. This could even lead to a
stabilization of the dilaton, so that one cannot extrapolate between weak and
strong coupling anymore. Obviously, the explicit knowledge of supersymmetric
black hole solutions in AdS is a necessary ingredient if one wants to study this
new scenario.

A first step in this direction was made in \cite{Cacciatori:2009iz,Klemm:2011xw}, where the
first examples of extremal static or rotating BPS black holes in AdS$_4$ with nontrivial scalar
field profiles were constructed. Thereby, essential use was made of the results
of \cite{Cacciatori:2008ek}, where all supersymmetric backgrounds (with a timelike Killing
spinor) of ${\cal N}=2$, $D=4$ matter-coupled supergravity with Fayet-Iliopoulos gauging
were classified. This provides
a systematic method to obtain BPS solutions, without the necessity to
guess some suitable ans\"atze. Perhaps one of the most important results of \cite{Cacciatori:2009iz}
was the construction of genuine static supersymmetric black holes with spherical
symmetry in the stu model. A crucial ingredient for the existence of these solutions is the presence
of nonconstant scalar fields. These black holes were then
further studied and generalized in \cite{Hristov:2010ri,Dall'Agata:2010gj}.

In this paper, we shall go one step further with respect to \cite{Cacciatori:2009iz},
and include also nut charge. Apart from the supersymmetric Reissner-Nordstr\"om-Taub-Nut-AdS
family in minimal gauged supergravity \cite{AlonsoAlberca:2000cs}, there are, to the best of our
knowledge, no other known BPS solutions of this type. In addition to providing an interesting
scenario to study holography \cite{Chamblin:1998pz,Hawking:1998ct,Leigh:2011au}, these are 
intriguing for the following reason: In gauged supergravity, electric-magnetic duality invariance
is obviously broken due to the minimal coupling of the gravitinos to the vector potential
(unless one introduces also a magnetic gauging, but we shall not do this in what follows).
Nevertheless, it was discovered in \cite{AlonsoAlberca:2000cs} that supersymmetric solutions
of minimal gauged supergravity still enjoy a sort of electric-magnetic duality invariance in which
electric and magnetic charges and mass and nut charge are rotated simultaneously.
A deeper understanding of this mysterious duality might reveal unexpected geometric 
structures underlying gauged supergravity theories.

In addition to the motivation given above, a further reason for considering
supersymmetric nut-charged AdS black holes is the attractor mechanism \cite{Ferrara:1995ih,
Strominger:1996kf,Ferrara:1996dd,Ferrara:1996um,Ferrara:1997tw}, that has been the subject
of extensive research in the asymptotically flat case, but for which only little is known for
spacetimes that asymptote to AdS. First steps towards a systematic analysis of
the attractor flow in gauged supergravity were made in \cite{Morales:2006gm,Bellucci:2008cb} for the
non-BPS and in \cite{Huebscher:2007hj,Cacciatori:2009iz,Dall'Agata:2010gj,Kachru:2011ps}  for the
BPS case, but it would be interesting to generalize these results to
include also nut charge. In fact, what we shall find here is that (at least for the simple prepotential
considered below) the flat directions in the effective black hole potential (which generically occur
in the BPS flow in gauged supergravity \cite{Cacciatori:2009iz}) are lifted by turning on a nut charge.

The remainder of this paper is organized as follows: In the next section, we
briefly review ${\cal N}=2$, $D=4$ gauged supergravity coupled to abelian vector
multiplets (presence of U$(1)$ Fayet-Iliopoulos terms) and give the general recipe
to construct supersymmetric solutions found in \cite{Cacciatori:2008ek}. In \ref{construction},
the equations of \cite{Cacciatori:2008ek} are solved for  the $\text{SU}(1,1)/\text{U}(1)$ model
with prepotential $F=-iX^0X^1$, and a class of one-quarter BPS black holes carrying two electric, two
magnetic and one nut charge is constructed. We also discuss the attractor mechanism for this solution
and its near-horizon limit. Moreover, it is shown how the results of \cite{AlonsoAlberca:2000cs}
are recovered in the case of constant moduli. In section \ref{lifting}, we oxidize the solution to
eleven dimensions and comment on its M-theory interpretation. 
Section \ref{final} contains our conclusions and some final remarks.

\section{The supersymmetric backgrounds of ${\cal N}=2$, $D=4$ gauged supergravity}
\label{sugra}

We consider ${\cal N}=2$, $D=4$ gauged supergravity coupled to $n_V$ abelian
vector multiplets \cite{Andrianopoli:1996cm}\footnote{Throughout this paper,
we use the notations and conventions of \cite{Vambroes}.}.
Apart from the vierbein $e^a_{\mu}$, the bosonic field content includes the
vectors $A^I_{\mu}$ enumerated by $I=0,\ldots,n_V$, and the complex scalars
$z^{\alpha}$ where $\alpha=1,\ldots,n_V$. These scalars parametrize
a special K\"ahler manifold, i.~e.~, an $n_V$-dimensional
Hodge-K\"ahler manifold that is the base of a symplectic bundle, with the
covariantly holomorphic sections
\begin{equation}
{\cal V} = \left(\begin{array}{c} X^I \\ F_I\end{array}\right)\,, \qquad
{\cal D}_{\bar\alpha}{\cal V} = \partial_{\bar\alpha}{\cal V}-\frac 12
(\partial_{\bar\alpha}{\cal K}){\cal V}=0\,, \label{sympl-vec}
\end{equation}
where ${\cal K}$ is the K\"ahler potential and ${\cal D}$ denotes the
K\"ahler-covariant derivative. ${\cal V}$ obeys the symplectic constraint
\begin{equation}
\langle {\cal V}\,,\bar{\cal V}\rangle = X^I\bar F_I-F_I\bar X^I=i\,.
\end{equation}
To solve this condition, one defines
\begin{equation}
{\cal V}=e^{{\cal K}(z,\bar z)/2}v(z)\,,
\end{equation}
where $v(z)$ is a holomorphic symplectic vector,
\begin{equation}
v(z) = \left(\begin{array}{c} Z^I(z) \\ \frac{\partial}{\partial Z^I}F(Z)
\end{array}\right)\,.
\end{equation}
F is a homogeneous function of degree two, called the prepotential,
whose existence is assumed to obtain the last expression.
The K\"ahler potential is then
\begin{equation}
e^{-{\cal K}(z,\bar z)} = -i\langle v\,,\bar v\rangle\,.
\end{equation}
The matrix ${\cal N}_{IJ}$ determining the coupling between the scalars
$z^{\alpha}$ and the vectors $A^I_{\mu}$ is defined by the relations
\begin{equation}\label{defN}
F_I = {\cal N}_{IJ}X^J\,, \qquad {\cal D}_{\bar\alpha}\bar F_I = {\cal N}_{IJ}
{\cal D}_{\bar\alpha}\bar X^J\,.
\end{equation}
The bosonic action reads
\begin{eqnarray}
e^{-1}{\cal L}_{\text{bos}} &=& \frac 12R + \frac 14(\text{Im}\,
{\cal N})_{IJ}F^I_{\mu\nu}F^{J\mu\nu} - \frac 18(\text{Re}\,{\cal N})_{IJ}\,e^{-1}
\epsilon^{\mu\nu\rho\sigma}F^I_{\mu\nu}F^J_{\rho\sigma} \nonumber \\
&& -g_{\alpha\bar\beta}\partial_{\mu}z^{\alpha}\partial^{\mu}\bar z^{\bar\beta}
- V\,, \label{action}
\end{eqnarray}
with the scalar potential
\eq
V = -2g^2\xi_I\xi_J[(\text{Im}\,{\cal N})^{-1|IJ}+8\bar X^IX^J]\,,
\feq
that results from U$(1)$ Fayet-Iliopoulos gauging. Here, $g$ denotes the
gauge coupling and the $\xi_I$ are constants. In what follows, we define
$g_I=g\xi_I$.

The most general timelike supersymmetric background of the theory described
above was constructed in \cite{Cacciatori:2008ek}, and is given by
\eq
ds^2 = -4|b|^2(dt+\sigma)^2 + |b|^{-2}(dz^2+e^{2\Phi}dwd\bar w)\ , \label{gen-metr}
\feq
where the complex function $b(z,w,\bar w)$, the real function $\Phi(z,w,\bar w)$
and the one-form $\sigma=\sigma_wdw+\sigma_{\bar w}d\bar w$, together with the
symplectic section \eqref{sympl-vec}\footnote{Note that also $\sigma$ and
$\cal V$ are independent of $t$.} are determined by the equations
\eq
\partial_z\Phi = 2ig_I\left(\frac{{\bar X}^I}b-\frac{X^I}{\bar b}\right)\ ,
\label{dzPhi}
\feq
\begin{eqnarray}
&&\qquad 4\partial\bar\partial\left(\frac{X^I}{\bar b}-\frac{\bar X^I}b\right) + \partial_z\left[e^{2\Phi}\partial_z
\left(\frac{X^I}{\bar b}-\frac{\bar X^I}b\right)\right]  \label{bianchi} \\
&&-2ig_J\partial_z\left\{e^{2\Phi}\left[|b|^{-2}(\text{Im}\,{\cal N})^{-1|IJ}
+ 2\left(\frac{X^I}{\bar b}+\frac{\bar X^I}b\right)\left(\frac{X^J}{\bar b}+\frac{\bar X^J}b\right)\right]\right\}= 0\,,
\nonumber
\end{eqnarray}
\begin{eqnarray}
&&\qquad 4\partial\bar\partial\left(\frac{F_I}{\bar b}-\frac{\bar F_I}b\right) + \partial_z\left[e^{2\Phi}\partial_z
\left(\frac{F_I}{\bar b}-\frac{\bar F_I}b\right)\right] \nonumber \\
&&-2ig_J\partial_z\left\{e^{2\Phi}\left[|b|^{-2}\text{Re}\,{\cal N}_{IL}(\text{Im}\,{\cal N})^{-1|JL}
+ 2\left(\frac{F_I}{\bar b}+\frac{\bar F_I}b\right)\left(\frac{X^J}{\bar b}+\frac{\bar X^J}b\right)\right]\right\}
\nonumber \\
&&-8ig_I e^{2\Phi}\left[\langle {\cal I}\,,\partial_z {\cal I}\rangle-\frac{g_J}{|b|^2}\left(\frac{X^J}{\bar b}
+\frac{\bar X^J}b\right)\right] = 0\,, \label{maxwell}
\end{eqnarray}
\begin{equation}
2\partial\bar\partial\Phi=e^{2\Phi}\left[ig_I\partial_z\left(\frac{X^I}{\bar b}-\frac{\bar X^I}b\right)
+\frac2{|b|^2}g_Ig_J(\text{Im}\,{\cal N})^{-1|IJ}+4\left(\frac{g_I X^I}{\bar b}+\frac{g_I \bar X^I}b
\right)^2\right]\,, \label{Delta-Phi}
\end{equation}
\begin{equation}
d\sigma + 2\,\star^{(3)}\!\langle{\cal I}\,,d{\cal I}\rangle - \frac i{|b|^2}g_I\left(\frac{\bar X^I}b
+\frac{X^I}{\bar b}\right)e^{2\Phi}dw\wedge d\bar w=0\,. \label{dsigma}
\end{equation}
Here $\star^{(3)}$ is the Hodge star on the three-dimensional base with metric\footnote{Whereas
in the ungauged case, this base space is flat and thus has trivial holonomy, here we have U(1)
holonomy with torsion \cite{Cacciatori:2008ek}.}
\eq
ds_3^2 = dz^2+e^{2\Phi}dwd\bar w\ ,
\feq
and we defined $\partial=\partial_w$, $\bar\partial=\partial_{\bar w}$, as well as
\begin{equation}
{\cal I} = \text{Im}\left({\cal V}/\bar b\right)\ .
\end{equation}
Given $b$, $\Phi$, $\sigma$ and $\cal V$, the fluxes read
\begin{eqnarray}
F^I&=&2(dt+\sigma)\wedge d\left[bX^I+\bar b\bar X^I\right]+|b|^{-2}dz\wedge d\bar w
\left[\bar X^I(\bar\partial\bar b+iA_{\bar w}\bar b)+({\cal D}_{\alpha}X^I)b\bar\partial z^{\alpha}-
\right. \nonumber \\
&&\left. X^I(\bar\partial b-iA_{\bar w}b)-({\cal D}_{\bar\alpha}\bar X^I)\bar b\bar\partial\bar z^{\bar\alpha}
\right]-|b|^{-2}dz\wedge dw\left[\bar X^I(\partial\bar b+iA_w\bar b)+\right. \nonumber \\
&&\left.({\cal D}_{\alpha}X^I)b\partial z^{\alpha}-X^I(\partial b-iA_w b)-({\cal D}_{\bar\alpha}\bar X^I)
\bar b\partial\bar z^{\bar\alpha}\right]- \nonumber \\
&&\frac 12|b|^{-2}e^{2\Phi}dw\wedge d\bar w\left[\bar X^I(\partial_z\bar b+iA_z\bar b)+({\cal D}_{\alpha}
X^I)b\partial_z z^{\alpha}-X^I(\partial_z b-iA_z b)- \right.\nonumber \\
&&\left.({\cal D}_{\bar\alpha}\bar X^I)\bar b\partial_z\bar z^{\bar\alpha}-2ig_J
(\text{Im}\,{\cal N})^{-1|IJ}\right]\,. \label{fluxes}
\end{eqnarray}
In \eqref{fluxes}, $A_{\mu}$ is the gauge field of the K\"ahler U$(1)$,
\eq
A_{\mu} = -\frac i2(\partial_{\alpha}{\cal K}\partial_{\mu}z^{\alpha} -
         \partial_{\bar\alpha}{\cal K}\partial_{\mu}{\bar z}^{\bar\alpha})\,.
\feq

\section{Constructing nut-charged black holes}
\label{construction}

In this section we shall obtain supersymmetric nut-charged black holes,
which have nontrivial moduli turned on.
In order to solve the system \eqref{dzPhi}-\eqref{dsigma} we shall assume that both $z^{\alpha}$ and $b$ 
depend on the coordinate $z$ only, and use the separation ansatz $\Phi=\psi(z)+\gamma(w,\bar w)$. Then
\eqref{dzPhi} becomes
\begin{equation}
  \label{psi}
	\psi'=2i\left(\frac{\bar X}b-\frac{X}{\bar b}\right)\,,
\end{equation}
where a prime denotes differentiation with respect to $z$ and $X\equiv g_IX^I$. Furthermore, we can
integrate \eqref{bianchi} once, with the result
\begin{equation}
\label{Bint}
\begin{split}
e^{2\psi}\partial_{z}\left(\frac{X^{I}}{\bar b}-\frac{{\bar X}^I}b\right)-2ie^{2\psi}\left[|b|^{-2}
(\text{Im}\,{\cal N})^{-1|IJ}g_{J}+\right.\\
+\left.2\left(\frac{X^I}{\bar b}+\frac{{\bar X}^I}{b}\right)\left(\frac X{\bar b}+\frac{\bar X}b\right)\right] = -4\pi ip^I\,,
\end{split}
\end{equation}  
where $p^I$ are related to the magnetic charges, as we shall see later.
Using the contraction of (\ref{Bint}) with $g_I$, \eqref{Delta-Phi} boils down to
\begin{equation}
\label{Liou}
-4\partial\bar\partial\gamma=\kappa e^{2\gamma}\,, \qquad \kappa = -8\pi g_Ip^I\,.
\end{equation}
This is the Liouville equation, which implies that the metric $e^{2\gamma}dwd\bar w$ has constant curvature
$\kappa$, determined by the $p^I$.

\subsection{SU(1,1)/U(1) model}
\label{X^0X^1model}

In what follows we shall specialize to the SU(1,1)/U(1) model with prepotential $F=-iX^0X^1$, that has
$n_V=1$ (one vector multiplet), and thus just one complex scalar $\tau$. Choosing $Z^0=1$, $Z^1=\tau$,
the symplectic vector $v$ becomes
\eq
v = \left(\begin{array}{c} 1 \\ \tau \\ -i\tau \\ -i\end{array}\right)\ .
\label{v-X0X1}
\feq
The K\"ahler potential, metric and kinetic matrix for the vectors are given
respectively by
\eq
e^{-{\cal K}} = 2(\tau + \bar\tau)\ , \qquad g_{\tau\bar\tau} = \partial_\tau\partial_{\bar\tau}
{\cal K} = (\tau + \bar\tau)^{-2}\ ,
\feq
\eq
{\cal N} = \left(\begin{array}{cc} -i\tau & 0 \\ 0 & -\frac i\tau\end{array}\right)\ .
\feq
Note that positivity of the kinetic terms in the action requires
${\mathrm{Re}}\tau>0$. For the scalar potential one obtains
\eq
V = -\frac4{\tau+\bar\tau}(g_0^2 + 2g_0g_1\tau + 2g_0g_1\bar\tau
+ g_1^2\tau\bar\tau)\ , \label{pot_su11}
\feq
which has an extremum at $\tau=\bar\tau=|g_0/g_1|$. In what follows we assume
$g_I>0$. Notice also that $F_I=-i\eta_{IJ}X^J$, where
\eq
\eta_{IJ} = \begin{pmatrix}
0 & 1\\
1 & 0	
\end{pmatrix}\,.
\feq
Moreover, $(\text{Im}\,{\cal N})^{-1}=-4\,\text{diag}(|X^0|^2, |X^1|^2)$.	

For this model, \eqref{maxwell} becomes
\begin{equation}
\label{Max}
\begin{split}
\partial_z\left[e^{2\psi}(-2i\eta_{IJ})\partial_z\text{Re}\left(\frac{X^J}{\bar b}\right)\right]-2i\partial_z\left\{e^{2\psi}
\left[|b|^{-2}\text{Re}\,{\cal N}_{IL}\left(\text{Im}\,{\cal N}\right)^{-1|JL}g_J\right.\right.+ \\ +\left.\left.8\text{Re}
\left(\frac{F_I}{\bar b}\right)\text{Re}\left(\frac X{\bar b}\right)\right]\right\}-8ig_I e^{2\psi}\left[-\frac i2\alpha_{KJ}
\left(\frac{{\bar X}^K}b\partial_z\frac{X^J}{\bar b} - \frac{X^K}{\bar b}\partial_z\frac{{\bar X}^J}b\right)\right.+\\
\left.-|b|^{-2}\left(\frac X{\bar b}+\frac{\bar X}b\right)\right]=0\,.
\end{split}
\end{equation}
We now make the ansatz
\begin{equation}
\frac{X^I}{\bar b} = \frac{\alpha^I z+\beta^I}{Az^2+Bz+C}\,, \label{ans-Xb}
\end{equation}
where $A$, $B$, $C$, $\alpha^I$ and $\beta^I$ are complex constants\footnote{Note that \eqref{ans-Xb}
generalizes the ansatz used in \cite{Cacciatori:2009iz} to obtain black holes without nut charge.}.
Without loss of generality, we can take $A=1$ and $B=iD$, with $D\in\bR$, since we are free to shift
$z\mapsto z-\text{Re}B/2$. As a consequence, \eqref{psi} reduces to
\begin{equation}
\label{psi1}
\psi' = 4\,\frac{\text{Im}\alpha z^3+z^2\left(\text{Im}\beta-D\text{Re}\alpha\right)-z\left(\text{Im}\left(\bar\alpha C
\right)+D\text{Re}\beta\right)-\text{Im}(\bar\beta C)}{z^4+z^2\left(2\text{Re}C+D^2\right)+2Dz\text{Im}C+|C|^2}\,,
\end{equation}
with $\alpha\equiv g_I\alpha^I$, $\beta\equiv g_I\beta^I$.
Inspired by minimal gauged supergravity \cite{Caldarelli:2003pb}, we choose
\begin{equation}
\label{1con}
\text{Im}\beta-D\text{Re}\alpha=0\,,
\end{equation}
\begin{equation}
\label{2con}
-4\left(\text{Im}\left(\bar\alpha C\right)+D\text{Re}\beta\right)=2\text{Im}\alpha\left(2\text{Re}C+D^2\right)\,,
\end{equation}
\begin{equation}
\label{3con}
-4\text{Im}\left(\bar\beta C\right)=2D\text{Im}\alpha\text{Im}C\,,
\end{equation} 
so that (\ref{psi1}) simplifies to
\begin{equation}
\psi' = \frac{4z^3+2(2\text{Re}C+D^2)z+2D\text{Im}C}{z^4+z^2\left(2\text{Re}C+D^2\right)+2Dz\text{Im}C+|C|^2}
\text{Im}\alpha\,,
\end{equation}
which can be integrated once to give
\begin{equation}
\psi = \text{Im}\alpha\left(\ln\left[z^4+z^2\left(2\text{Re}C+D^2\right)+2Dz\text{Im}C+|C|^2\right]+\ln\check C
\right)\,,
\end{equation}
where $\check C$ denotes an integration constant that can be set to $1$ without loss of generality
by using the scaling symmetry $\psi\mapsto\psi-\ln\lambda$, $\gamma\mapsto\gamma+\ln\lambda$,
$\kappa\mapsto\kappa/\lambda^2$, $p^I\mapsto p^I/\lambda^2$, with $\ln\lambda=\text{Im}\alpha\ln\check C$, 
that leaves \eqref{psi}, \eqref{Bint} and \eqref{Liou} invariant.

In order to solve \eqref{Bint}, we take into account that
\begin{displaymath}
|b|^{-2}(\text{Im}\,{\cal N})^{-1|IJ}g_J = -4\frac{|X^I|^2}{|b|^2}g_I\,,
\end{displaymath}
where there is of course no summation over $I$ on the rhs. Then \eqref{Bint} becomes
\begin{eqnarray}
\label{Bpol}
-4\pi ip^{I}=\left[(z^2+iDz+C)\left(z^2-iDz+\bar C\right)\right]^{2\text{Im}\alpha}\times\quad &&\nonumber \\
\times\left\{\left[\frac{-\alpha^Iz^2-2\beta^Iz+\alpha^IC-\beta^IiD}{\left(z^2+iDz+C\right)^2}+\frac{\bar \alpha^Iz^2
+2\bar\beta^Iz-\bar\alpha^I\bar C-\bar\beta^IiD}{\left(z^2-iDz+\bar C\right)^2}\right]\right.\quad &&\nonumber \\
\left.-2i\left[-4g_I\left|\frac{\alpha^Iz+\beta^I}{z^2+iDz+C}\right|^2+8\text{Re}\left(\frac{\alpha^Iz+\beta^I}
{z^2+iDz+C}\right)\text{Re}\left(\frac{\alpha z+\beta}{z^2+iDz+C}\right)\right]\right\}.\quad &&
\end{eqnarray}
In order to simplify the calculations further, we shall also take $\text{Im}\alpha=1/2$, so that
\eqref{Bpol} boils down to a sixth order polynomial equation,
\begin{equation}
A_0 + A_1z + A_2z^2 + A_3z^3 + A_4z^4 + A_5z^5 + A_6z^6 = 0\,, \label{6thorder}
\end{equation}
where $A_6=0$ iff
\begin{equation}
-\text{Im}\alpha^I+4g_I|\alpha^I|^2-8\text{Re}\alpha\,\text{Re}\alpha^I=0\,,
\end{equation}
and thus
\begin{equation} \label{ImalphaI}
\text{Im}\alpha^I=\frac{1\pm\sqrt{1-16g_I(4g_I\text{Re}^2\alpha^I-8\text{Re}\alpha\,\text{Re}\alpha^I)}}{8g_I}\,.
\end{equation}
Using $\text{Im}\alpha=1/2$, and defining $8g_0\text{Re}\alpha^0\equiv x$, $8g_1\text{Re}\alpha^1\equiv y$, this yields
\begin{equation}
\label{poli}
x^4+y^4-8(x^2+y^2)-2x^2y^2-32xy=0\,.
\end{equation}
To proceed further, recall that
\eq
\label{scalar}
\tau = \frac{Z^1}{Z^0} = \frac{X^1}{X^0} = \frac{\alpha^1z+\beta^1}{\alpha^0z+\beta^0}\,.
\feq
If we require that the scalar asymptotically approaches the AdS vacuum, that is $\tau\to g_0/g_1$
for $z\to\infty$, we must have $\alpha^1/\alpha^0=g_0/g_1$, and thus $x=y$. \eqref{poli} implies then
$x=0$, hence $\text{Re}\alpha^I=0$. Plugging this into \eqref{ImalphaI} gives\footnote{Taking the lower sign
yields $\text{Im}\alpha^I=0$, and thus a constant scalar.}
\begin{equation}
\text{Im}\alpha^I = \frac1{4g_I}\,.
\end{equation}
\eqref{1con} and \eqref{2con} reduce respectively to
\eq
\text{Im}\beta = 0\,, \qquad \text{Re}\beta = -\frac D4\,, \label{Re-beta}
\feq
implying
\begin{equation}
\label{const2}
\beta = -\frac D4\,.
\end{equation}
Using the above results, one finds that \eqref{3con} is identically satisfied.

Let us go back to \eqref{6thorder}. Requiring $A_0=0$ leads to
\begin{equation}
\label{magcharge}
-4\pi p^I = \frac{\text{Re}C}{2g_I}+8g_I|\beta^I|^2+2D\text{Re}\beta^I\,,
\end{equation}
which gives the magnetic charges in terms of some numerical constants.
Note that the above equation, together with \eqref{const2}, implies
\eq
g_0p^0 = g_1p^1\,, \qquad \kappa = -16\pi g_0p^0 = -16\pi g_1p^1\,.
\feq
Eventually, one finds that $A_6=0$ and $A_0=0$ are sufficient conditions for \eqref{6thorder} to be satisfied.

We now turn to \eqref{maxwell}. After some lengthy calculations, one gets
\begin{equation}
\label{lastMax}
e^{2\psi}\left[\left\langle\mathcal{I},\partial_z\mathcal{I}\right\rangle-|b|^{-2}\left(\frac{X}{\bar b}+\frac{\bar X}b
\right)\right] = -\frac D{16g_0g_1}\,,
\end{equation}
and thus \eqref{maxwell} can be integrated once to give
\begin{equation}
\label{Maxint}
\begin{split}
e^{2\psi}\partial_z\left[2i\text{Im}\left(\frac{F_I}{\bar b}\right)\right]-2ig_Je^{2\psi}\left[|b|^{-2}\text{Re}\,
{\cal N}_{IL}\left(\text{Im}\,{\cal N}\right)^{-1|JL}\right.+\\
+\left.8\text{Re}\left(\frac{F_I}{\bar b}\right)\text{Re}\left(\frac{X^J}{\bar b}\right)\right]+i\frac{g_ID}{2g_0g_1}z
= -4\pi i q_I\,,	
\end{split}
\end{equation}
where $q_I$ are related to the electric charges. In order to solve \eqref{Maxint}, notice that
\begin{equation}
\text{Re}\,{\cal N} = \frac{{\bar X}^0X^1-{\bar X}^1X^0}{2i}
\begin{pmatrix}
|X^0|^{-2} & 0\\
0 & -|X^1|^{-2}	
\end{pmatrix}\,,	
\end{equation}
from which
\begin{displaymath}
\text{Re}\,{\cal N}_{IL}\left(\text{Im}\,{\cal N}\right)^{-1|JL}g_J = 2i\left({\bar X}^0X^1-{\bar X}^1X^0\right)(-1)^Ig_I
\qquad \text{(no summation over $I$).}
\end{displaymath}
Using this, \eqref{Maxint} boils down to a fifth order polynomial equation,
\begin{equation}
B_0+B_1z+B_2z^2+B_3z^3+B_4z^4+B_5z^5 = 0\,. \label{5thorder}
\end{equation}
One finds that $B_5$ vanishes identically provided that \eqref{Re-beta} holds.
Requiring $B_0=0$ yields
\eq
\label{qch}
\eta_{IJ}\left(\frac{\text{Im}C}{4g_J}-D\text{Im}\beta^J\right)+4(-1)^Ig_I\text{Im}(\beta^1{\bar\beta}^0)=2\pi q_I\,,
\feq
which determines the electric charges. Given that \eqref{const2} holds, the above equation leads to
\begin{equation}
\label{imc}
g_I^{-1}\text{Im}C = 8\left(\pi\eta^{IJ}q_J + \text{Im}\beta^ID\right)\,, 
\end{equation}
where $\eta^{IJ}$ denotes the inverse of $\eta_{IJ}$. Note that \eqref{qch} implies also
\begin{equation}
\text{Im}C = 4\pi(g_1q_0+g_0q_1)\,,
\end{equation}
which, combined with (\ref{imc}), yields (no summation over $I$)
\begin{equation}
(-1)^Ig_I\text{Im}\beta^I = \frac{\pi\left(g_1q_0 - g_0q_1\right)}{2D}\,. 
\end{equation}
It turns out that then all coefficients in \eqref{5thorder} vanish, and thus \eqref{Maxint} is satisfied.

Finally, taking
\eq
e^{2\gamma} = \left(1+\frac{\kappa}4 w\bar w\right)^{-2}
\feq
as a solution of the Liouville equation \eqref{Liou}, one can compute the shift vector from \eqref{dsigma},
with the result
\eq
\sigma = \frac{iD}{32g_0g_1}\frac{w d\bar w - \bar w dw}{1+\frac{\kappa}4 w\bar w}\,.
\feq
Note that $d\sigma$ is proportional to the K\"ahler form on the two-space with metric
$e^{2\gamma}dwd\bar w$. The four-dimensional line element reads
\begin{equation}
\label{fmetric}
ds^2 = -4|b|^2(dt+\sigma)^2+\frac{dz^2}{|b|^2}+\frac{z^2+16g_0g_1\text{Re}(\beta^1\bar\beta^0)}{4g_0g_1}
\frac{dwd\bar w}{\left(1+\frac{\kappa}4 w\bar w\right)^2}\,,
\end{equation}
where
\begin{equation}
\label{b}
|b|^2 = 4g_0g_1\frac{|z^2+iDz+C|^2}{z^2+16g_0g_1\text{Re}(\beta^1\bar\beta^0)}\,.
\end{equation}
As we said, positivity of the kinetic terms in the action requires $\text{Re}\tau>0$. From \eqref{scalar}
one sees that this is equivalent to
\begin{equation}
z^2 > -16g_0g_1\text{Re}(\beta^1\bar\beta^0)\,.
\end{equation}
As can be seen from \eqref{b}, $|b|$ diverges when $\text{Re}\tau=0$, signaling the presence of a
curvature singularity at the point where ghost  modes appear.
The solution we have found will have an event horizon for some $z=z_{\text h}$, with
$z_{\text h}^2+iDz_{\text h}+C=0$, and thus $z_{\text h}^2=-\text{Re}C$ and $Dz_{\text h}=-\text{Im}C$, which
in turn imply
\eq
\text{Im}^2 C = -D^2\text{Re}C\,, \label{constr-hor}
\feq
and therefore $\text{Re}C<0$. Putting these results together, we can be more specific about the geometry
of the horizon. First of all, contracting \eqref{magcharge} with $g_I$ and taking into account \eqref{const2}
and the second equation of \eqref{Liou} yields
\begin{equation}
\label{kappa}
\kappa = 2\text{Re}C+16\sum_Ig_I^2|\beta^I|^2-D^2\,.
\end{equation}
If we want the dangerous point where ghost modes appear to be hidden behind the horizon, we must have
$-\text{Re}C>-16g_0g_1\text{Re}(\beta^1\bar\beta^0)$. Using this in \eqref{kappa} gives
\begin{equation}
\kappa < 16|\beta|^2-D^2\,,
\end{equation}
which, together with \eqref{const2}, yields $\kappa<0$, so that the horizon must be hyperbolic.   
Note that one can also have solutions with spherical instead of hyperbolic symmetry, but these
are naked singularities. A special case occurs for $\kappa=0$, i.e., for a flat horizon. Then, the
point where ghost modes appear coincides with the horizon. The resulting geometry describes a Nernst
brane \cite{Barisch:2011ui}, whose entropy vanishes at zero temperature. Solutions of this type have
potential applications in AdS/cond-mat, but unfortunately for $\kappa=0$ the spacetime \eqref{fmetric}
has a curvature singularity at $z=z_{\text h}$, where
\eq
R_{\mu\nu\rho\sigma}R^{\mu\nu\rho\sigma} \sim (z-z_{\text h})^{-2}\,.
\feq
Coming back to the case of arbitrary $\kappa$,
the fluxes can be computed from \eqref{fluxes}, with the result (no summation over $I$)
\begin{eqnarray}
\label{fluxes-nut}
F^I &=& (dt+\sigma)\wedge dz\frac{16g_0g_1}{(z^2+16g_0g_1\text{Re}(\beta^0\bar\beta^1))^2}\left[\left(
\frac{\text{Im}C}{4g_I}+D\text{Im}\beta^I\right)(16g_0g_1\text{Re}(\beta^0\bar\beta^1)-z^2)\right. \nonumber \\
&&+\left.2z\left(16g_0g_1\text{Re}(\beta^0\bar\beta^1)\left(\text{Re}\beta^I+\frac{D}{4g_I}\right)-\text{Re}(\beta^I
\bar C)\right)\right]-\frac{ie^{2\gamma}dw\wedge d\bar w}{z^2+16g_0g_1\text{Re}(\beta^0\bar\beta^1)}
\cdot \nonumber \\
&&\cdot\left[\left(\frac{D^2}{4g_I}+D\text{Re}\beta^I+\frac{\kappa}{8g_I}\right)z^2
+D\left(\frac{\text{Im}C}{4g_I}+D\text{Im}\beta^I\right)z+D\text{Re}(\beta^I\bar C)\right. \nonumber \\
&&+\left.\frac{\kappa}{8g_I}\left(\text{Re}C-\frac{\kappa}2\right)\right]\,.
\end{eqnarray}
To sum up, the metric is given by \eqref{fmetric}, the U(1) field strengths by \eqref{fluxes-nut}, and the
complex scalar $\tau$ reads
\eq
\tau = \frac{g_0}{g_1}\frac{z-4ig_1\beta^1}{z-4ig_0\beta^0}\,, \label{tau-final}
\feq
where the constants $\beta^I\in\bC$ are constrained by \eqref{const2}. A priori, the solution is labelled
by the 7 real parameters $\beta^I,C,D$, but \eqref{const2}, together with \eqref{kappa}, leave 4 independent
constants. Note that $\kappa$ can be set to $0,\pm 1$ without loss of generality by using the scaling
symmetry $(t,z,w,C,D,\beta^I,\kappa)\mapsto(t/\lambda,\lambda z,w/\lambda,\lambda^2C,\lambda D,
\lambda\beta^I,\lambda^2\kappa)$ leaving the metric, fluxes and scalar invariant. A convenient way
of parametrizing the constraint \eqref{const2} is
\eq
g_0\text{Im}\beta^0 = -g_1\text{Im}\beta^1 = \frac{\nu}4\,, \qquad g_0\text{Re}\beta^0 = \frac{\mu-n}4\,,
\qquad g_1\text{Re}\beta^1=-\frac{\mu+n}4\,, \label{parametr}
\feq
where $n=D/2$. Then, \eqref{b} becomes
\eq
|b|^2 = 4g_0g_1\frac{|z^2+2inz+C|^2}{z^2-\mu^2-\nu^2+n^2}\,,
\feq
with
\eq
\text{Re}C = \frac{\kappa}2-\mu^2-\nu^2+n^2\,.
\feq
If, in addition, we
want the metric to have a horizon, the additional constraint \eqref{constr-hor} must be satisfied. We have
thus obtained a three-parameter family ($\mu,\nu,n$) of black holes, whose nut charge is given by $n$.

The magnetic and electric charges read respectively
\begin{eqnarray}
P^I &=& \frac1{4\pi}\int_{\Sigma_{\infty}} F^I = \left[p^I - \frac{D^2}{8\pi g_I} - \frac D{2\pi}\text{Re}
\beta^I\right]V\,, \nonumber \\
Q_I &=& \frac1{4\pi}\int_{\Sigma_{\infty}} G_I = \left[q_I + \frac D{2\pi}\eta_{IJ}\text{Im}\beta^J\right]V\,,
\end{eqnarray}
where $G_{+I}={\cal N}_{IJ}F^{+J}$ \cite{Vambroes}, $\Sigma_{\infty}$ denotes a surface of constant
$t,z$ for $z\to\infty$, and $V$ is defined by
\begin{equation}
V=\frac i2\int{e^{2\gamma}dw\wedge d\bar w}\,.
\end{equation}
For $\kappa=-1$, this yields in terms of the parameters $\mu,\nu,n$
\begin{eqnarray}
P^0 &=& -\frac V{4\pi g_0}\left[n^2 + n\mu - \frac14\right]\,, \qquad P^1 = -\frac V{4\pi g_1}\left[n^2 -
n\mu - \frac14\right]\,, \\
Q_0 &=& \frac{nV}{4\pi g_1}\left[\sqrt{\frac12+\mu^2+\nu^2-n^2}+\nu\right]\,, \qquad
Q_1 = \frac{nV}{4\pi g_0}\left[\sqrt{\frac12+\mu^2+\nu^2-n^2}-\nu\right]\,. \nonumber
\end{eqnarray}
The value of the scalar field on the horizon and the entropy are
\eq
\tau_{\text h}=\frac{g_0}{g_1}\frac{\sqrt{\frac12+\mu^2+\nu^2-n^2}-\nu+i(\mu+n)}{\sqrt{\frac12+\mu^2+\nu^2
-n^2}+\nu-i(\mu-n)}\,, \qquad S=\frac{A_{\text h}}{4G}=\frac{\pi V}{4g_0g_1}\,, \label{entropy}
\feq
where we have taken into account that $8\pi G=1$ in our conventions. If the horizon is compactified
to a Riemann surface of genus $h>1$, we can use Gauss-Bonnet to get $V=4\pi(h-1)$, and thus
\eq
S = \frac{\pi^2(h-1)}{g_0g_1}\,. \label{S-compact}
\feq
For a noncompact horizon, $V$ is infinite, but the entropy- and charge densities are finite.
If we define the complex charge
\eq
z^I = P^I+i\eta^{IJ}Q_J\,,
\feq
as well as the symplectic vector
\eq
{\cal Z} = \left(\begin{array}{c} z^I \\ -i\eta_{IJ}z^J\end{array}\right)\,,
\feq
the Bekenstein-Hawking entropy can be rewritten in the form
\eq
S = -\frac{16i\pi^3}V\langle{\cal Z},\bar{\cal Z}\rangle\,,
\feq
where $\langle\cdot,\cdot\rangle$ denotes the symplectic product. For nonvanishing nut
parameter $n$, one can express $\tau_{\text h}$ in terms of the charges,
\eq
\tau_{\text h} = \frac{g_0}{g_1}\frac{1-16\pi g_0z^0/V}{1-16\pi g_1z^1/V}\,. \label{tau_h-charges}
\feq
If the nut charge is zero, both the nominator and the denominator of \eqref{tau_h-charges} vanish,
and $\tau_{\text h}$ ceases to be a function of the charges: In this case we have $Q_I=0$,
$P^I=V/(16\pi g_I)$, while $\tau_{\text h}$ depends on the two parameters $\mu,\nu$ which are
independent of the charges. The scalar field is thus not stabilized for $n=0$; $\tau_{\text h}$ takes
values in the moduli space $\text{SU}(1,1)/\text{U}(1)$\footnote{Nevertheless, the entropy is independent
of the values of the moduli on the horizon not fixed by the charges, in agreement with the attractor
mechanism \cite{Ferrara:1995ih,Strominger:1996kf,Ferrara:1996dd,Ferrara:1996um,Ferrara:1997tw}.}.
These flat directions are lifted by turning
on a nut parameter, since then $\tau_{\text h}$ is completely fixed by the charges, cf.~\eqref{tau_h-charges}.

\subsection{Near-horizon limit}

The near-horizon limit is obtained by setting $z=z_{\text h}+\epsilon\hat z$, $t=\hat t/(2\epsilon)$,
and taking the limit $\epsilon\to 0$. Then the metric \eqref{fmetric} boils down to
\eq
ds^2 = -\frac{\hat z^2}{L^2}d\hat t^2 + L^2\frac{d\hat z^2}{\hat z^2} + \frac{e^{2\gamma}dwd\bar w}
{8g_0g_1}\,,
\feq
which is $\text{AdS}_2\times\text{H}^2$, with the AdS length scale $L$ set by
\[
L^{-2} = 16g_0g_1(1+2\mu^2+2\nu^2)\,.
\]
Note that the shift vector $\sigma$ is scaled away in this limit. The near-horizon limit of the fluxes
\eqref{fluxes-nut} can be cast into the form
\eq
F^I = -8\text{Im}(X^I\bar X^Jg_J)d\hat t\wedge d\hat z + 2\pi ip^Ie^{2\gamma}dw\wedge d\bar w\,.
\feq

\subsection{Constant scalars}
\label{sec:const}

In order to shed further light on the physical meaning of the parameters appearing in \eqref{fmetric},
and to compare with the results of \cite{AlonsoAlberca:2000cs}, we will now consider the case of constant
scalars. As we are interested in solutions with genuine horizons, we take $\kappa=-1$ in what follows.

First of all, from \eqref{tau-final} it is clear that $\tau$ is constant iff $g_0\beta^0=g_1\beta^1$. Taking into
account \eqref{const2}, this implies
\eq
\beta^I = -\frac D{8g_I}\,. \label{betaI-const}
\feq
Since $\tau=g_0/g_1$, the scalar potential $V$ in \eqref{action} reduces to a cosmological constant
$\Lambda=-3/l^2$, with $l^{-2}=4g_0g_1$. Setting
\eq
z= \frac rl\,, \qquad w = 2e^{i\phi}\tanh\frac{\theta}2\,,
\feq
as well as\footnote{Notice that, with \eqref{betaI-const} and \eqref{def-MN}, the constraint \eqref{kappa} is
automatically satisfied.}
\eq
\text{Re}C = \frac{N^2}{l^2}-\frac12\,, \qquad \text{Im}C = -\frac M{2N}\,, \qquad D = \frac{2N}l\,,
\label{def-MN}
\feq
and transforming the time coordinate according to $t \mapsto l(N\phi - t/2)$, the metric \eqref{fmetric}
becomes
\eq
ds^2 = -\frac{\lambda}{r^2+N^2}(dt-2N\cosh\theta d\phi)^2 + \frac{r^2+N^2}{\lambda}dr^2 +
(r^2+N^2)(d\theta^2 + \sinh^2\theta d\phi^2)\,,
\feq
with $\lambda$ given by
\eq
\lambda = \frac1{l^2}(r^2+N^2)^2 + \left(-1+\frac{4N^2}{l^2}\right)(r^2-N^2) - 2Mr + \left(\frac{2N^2}l-\frac l2
\right)^2 + \frac{l^2M^2}{4N^2}\,. \label{lambda}
\feq
This represents a subclass of the (hyperbolic) Reissner-Nordstr\"om-Taub-NUT-AdS spacetime.
The fluxes \eqref{fluxes-nut} boil down to
\begin{eqnarray}
F^I &=& -(dt-2N\cosh\theta d\phi)\wedge\frac{dr}{2lg_I(r^2+N^2)^2}\left[\frac{Ml}{2N}(r^2-N^2) +
2rN\left(\frac{2N^2}l-\frac l2\right)\right] \nonumber \\
&& -\frac{\sinh\theta d\theta\wedge d\phi}{2lg_I(r^2+N^2)}\left[\left(\frac{2N^2}l-\frac l2\right)(r^2-N^2)
-Mlr\right]\,. \label{fluxes-const}
\end{eqnarray}
It is not difficult to see that the action \eqref{action} reduces to the one of minimal gauged supergravity
considered in \cite{AlonsoAlberca:2000cs} for $g_0F^0=g_1F^1\equiv F/(2l)$. The field strength $F$
computed this way from \eqref{fluxes-const} coincides exactly with the expression following from
the RN-TN-AdS gauge potential (2.4) of \cite{AlonsoAlberca:2000cs} if we identify
\eq
Q = -\frac{Ml}{2N}\,, \qquad P = \frac{2N^2}l - \frac l2\,. \label{QP}
\feq
These are precisely the conditions on the electric and magnetic charge found
in \cite{AlonsoAlberca:2000cs}, for which the
Reissner-Nordstr\"om-Taub-NUT-AdS solution is supersymmetric\footnote{Actually, the conditions
given in \cite{AlonsoAlberca:2000cs} are $Q=\mp Ml/(2N)$ and $P=\pm(2N^2/l-l/2)$, corresponding
to vanishing ${\cal B}_{\mp}$ in (3.10), (3.12) of \cite{AlonsoAlberca:2000cs}. We have here the upper
sign, but the lower one can easily be generated by the CPT transformation $\phi\mapsto-\phi$,
$t\mapsto-t$ (that leaves the metric invariant).}. Moreover, if \eqref{QP} holds, the function
\eqref{lambda} reduces to equ.~(2.1) of \cite{AlonsoAlberca:2000cs}.
As a nontrivial consistency check, we have thus reproduced the
known BPS conditions of minimal gauged supergravity. As we said, in order to have a horizon, the
additional constraint \eqref{constr-hor} must be satisfied. In this case, \eqref{constr-hor} leads to
\eq
M = \frac{4N^2}l\left(\frac12 - \frac{N^2}{l^2}\right)\,.
\feq
This leaves a one-parameter family of supersymmetric black holes, labelled by the nut charge $N$.
From \eqref{def-MN} it is also clear that the imaginary part of $C$ is related to the black hole mass.

\section{Lifting to M-theory}
\label{lifting}

We now want to uplift the black hole solutions obtained in section \ref{X^0X^1model} to M-theory, and
comment on their higher-dimensional interpretation. The Kaluza-Klein ansatz given in \cite{Cvetic:1999au}
allows to reduce eleven-dimensional supergravity to ${\cal N} = 4$ SO$(4)$ gauged supergravity in four dimensions, which can be further truncated to the $F=-iX^0X^1$ model of section \ref{X^0X^1model}.
The reduction ansatz for the metric reads \cite{Cvetic:1999au}
\eq
ds^2_{11} = \Delta^{2/3}ds^2_4 + \frac{2\Delta^{2/3}}{g^2}d\xi^2 + \frac{\Delta^{2/3}}{2g^2}\left[\frac{c^2}
{c^2X^2+s^2}\sum_{i=1}^3(h^i)^2 + \frac{s^2}{s^2\tilde X^2+c^2}\sum_{i=1}^3(\tilde h^i)^2\right]\,,
\label{metr11d}
\feq
where
\begin{eqnarray}
\tilde X &=& X^{-1}q\,, \qquad q^2 = 1+\chi^2 X^4\,, \qquad \Delta = \left[(c^2X^2+s^2)(s^2\tilde X^2+c^2)
\right]^{1/2}\,, \nonumber \\
c &=& \cos\xi\,, \qquad s = \sin\xi\,, \qquad h^i = \sigma_i - gA^i_{(1)}\,, \qquad \tilde h^i = \tilde\sigma_i
-g\tilde A^i_{(1)}\,.
\end{eqnarray}
Here, the $\sigma_i$ are left-invariant 1-forms on $\text{S}^3=\text{SU}(2)$, and $\tilde\sigma_i$ are
left-invariant 1-forms on a second S$^3$. They satisfy
\eq
d\sigma_i = -\frac12\epsilon_{ijk}\sigma_j\wedge\sigma_k\,, \qquad d\tilde\sigma_i = -\frac12\epsilon_{ijk}
\tilde\sigma_j\wedge\tilde\sigma_k\,. 
\feq
$A^i_{(1)},\tilde A^i_{(1)}$ denote the $\text{su}(2)\times\text{su}(2)\cong\text{so}(4)$ Yang-Mills potentials,
$g$ is the gauge coupling constant, and $X=\exp(\phi/2)$. $\phi$ and $\chi$ are the dilaton and axion
of the ${\cal N}=4$, $D=4$ theory respectively. The ansatz for the 4-form is given by \cite{Cvetic:1999au}
\eq
F_{(4)} =-g\sqrt2U\epsilon_{(4)} - \frac{4sc}{g\sqrt2}X^{-1}\ast\!dX\wedge d\xi + \frac{\sqrt2 sc}g\chi X^4
\ast\!d\chi\wedge d\xi + F'_{(4)} + F''_{(4)}\,,
\feq
with $\ast$ the Hodge dual operator of $ds_4^2$, and $\epsilon_{(4)}$ the corresponding volume form.
The expressions for $F'_{(4)}$ and $F''_{(4)}$ are rather lengthy, and can be found in eqns.~(9) and (10)
of \cite{Cvetic:1999au}. $U$ is defined by
\eq
U = X^2c^2+\tilde X^2s^2+2\,.
\feq
Plugging the above reduction ans\"atze into the eleven-dimensional equations of
motion gives rise to the equations of motion of ${\cal N}=4$, $D=4$ gauged supergravity. If we truncate
further by setting $A^1_{(1)}=A^2_{(1)}=\tilde A^1_{(1)}=\tilde A^2_{(1)}=0$ (which corresponds to
considering only the Cartan subgroup $\text{U}(1)\times\text{U}(1)$ of $\text{SO}(4)$), the bosonic
Lagrangian in four dimensions becomes \cite{Cvetic:1999au}
\begin{eqnarray}
{\cal L}_4 &=& R\ast\!1 - \frac12\ast\!d\phi\wedge d\phi - \frac12e^{2\phi}\ast\!d\chi\wedge d\chi - V\ast\!1
-\frac12 e^{-\phi}\ast\!F^3_{(2)}\wedge F^3_{(2)} \nonumber \\
&& -\frac12\frac{e^{\phi}}{1+\chi^2e^{2\phi}}\ast\!\tilde F^3_{(2)}\wedge\tilde F^3_{(2)} - \frac12\chi
F^3_{(2)}\wedge F^3_{(2)} + \frac12\frac{\chi e^{2\phi}}{1+\chi^2e^{2\phi}}\tilde F^3_{(2)}\wedge
\tilde F^3_{(2)}\,,
\end{eqnarray}
where $F^3_{(2)}=dA^3_{(1)},\tilde F^3_{(2)}=d\tilde A^3_{(1)}$, and the scalar potential reads
\eq
V = -2g^2(4 + 2\cosh\phi + \chi^2 e^{\phi})\,.
\feq
This is (up to a constant prefactor) equal to the Lagrangian \eqref{action} for the prepotential $F=-iX^0X^1$,
if we identify
\eq
F^0 = \frac1{\sqrt2}F^3_{(2)}\,, \qquad F^1 = \frac1{\sqrt2}\tilde F^3_{(2)}\,, \qquad\tau = e^{-\phi}-i\chi\,,
\feq
and take $g_0=g_1=g/\sqrt2$ for the gauge coupling constants. This allows to oxidize the solution
\eqref{fmetric}, \eqref{fluxes-nut}, \eqref{tau-final} to eleven dimensions. The functions $X,\tilde X$
are then given by
\eq
X^2 = \frac{(z+\nu)^2+(n-\mu)^2}{z^2-\mu^2-\nu^2+n^2}\,, \qquad
\tilde X^2 = \frac{(z-\nu)^2+(n+\mu)^2}{z^2-\mu^2-\nu^2+n^2}\,.
\feq
Choosing Euler angles $\psi,\vartheta,\varphi$ on the first $\text{S}^3$ and $\Psi,\Theta,\Phi$ on
the second $\text{S}^3$, we have for the left-invariant 1-forms
\begin{eqnarray*}
\sigma_1 &=& \sin\psi d\vartheta-\cos\psi\sin\vartheta d\varphi\,, \\
\sigma_2 &=& \cos\psi d\vartheta+\sin\psi\sin\vartheta d\varphi\,, \\
\sigma_3 &=& d\psi+\cos\vartheta d\varphi\,,
\end{eqnarray*}
and similar for $\tilde\sigma_i$. After that, the expressions $\sum_i(h^i)^2$ and $\sum_i(\tilde h^i)^2$ in
\eqref{metr11d} simplify in our case to
\begin{eqnarray}
\sum_{i=1}^3(h^i)^2 &=& d\vartheta^2 + \sin^2\vartheta d\varphi^2 + (d\psi+\cos\vartheta d\varphi
-gA^3_{(1)})^2\,, \\
\sum_{i=1}^3(\tilde h^i)^2 &=& d\Theta^2 + \sin^2\Theta d\Phi^2 + (d\Psi+\cos\Theta d\Phi - g\tilde A^3_{(1)})^2\,,
\nonumber
\end{eqnarray}
where
\begin{eqnarray*}
A^3_{(1)t} &=& 2g\frac{\kappa(n-\mu)+4n(-\mu^2-\nu^2+n^2)-2z(\text{Im}C+2n\nu)}{z^2-\mu^2-\nu^2+n^2}\,, \\
A^3_{(1)w} &=& -\frac{i\bar w}{2g(1+\frac{\kappa}4w\bar w)(z^2-\mu^2-\nu^2+n^2)}\left[\left(n^2+n\mu+
\frac{\kappa}4\right)z^2+nz(2n\nu+\text{Im}C)\right. \\
&& +\left.\left(\frac{\kappa}2-\mu^2-\nu^2+n^2\right)\left(\frac{\kappa}4-n^2+n\mu\right)+n\nu\text{Im}C-
\frac{\kappa^2}8\right]\,,
\end{eqnarray*}
$A^3_{(1)\bar w}=(A^3_{(1)w})^\star$ and $A^3_{(1)z}=0$. The expressions for $\tilde A^3_{(1)}$ result
from those for $A^3_{(1)}$ by replacing $\mu\to-\mu$ and $\nu\to-\nu$.

For $\mu=\nu=n=0$, the solution \eqref{metr11d} can be interpreted as the gravity dual corresponding
to membranes wrapping holomorphic curves in a Calabi-Yau five-fold \cite{Gauntlett:2001qs}. It
would be interesting to see whether the general solution \eqref{metr11d}
(for $\mu,\nu,n\neq 0$) has a similar interpretation.
This might allow for a microscopic entropy computation of the four-dimensional black hole
\eqref{fmetric}, which can then be compared with the macroscopic Bekenstein-Hawking result \eqref{S-compact}.

\section{Final remarks}
\label{final}

In this paper, we constructed a family of one-quarter BPS black holes in ${\cal N}=2$, $D=4$
FI-gauged supergravity carrying two electric,
two magnetic and one nut charge. The solution is given in terms of three free parameters, and
has a hyperbolic horizon. We saw that for vanishing nut charge, there are flat directions in
the effective black hole potential, in agreement with the results of \cite{Cacciatori:2009iz},
where a general near-horizon analysis was done. Turning on a nut parameter lifts these flat
directions, so that the horizon value of the moduli are completely fixed in terms of the charges.

A possible extension of our work would be to use the eleven-dimensional interpretation of our solution,
cf.~the oxidized metric obtained in section \ref{lifting}, to compute microscopically the entropy, which can
then be compared with the classical Bekenstein-Hawking result \eqref{entropy}. Moreover, it would be
interesting to consider other prepotentials, for instance the $t^3$ model, which allows for
supersymmetric black holes with spherical symmetry \cite{Cacciatori:2009iz}, and try to add rotation
and nut charge to the known static black holes \cite{Cacciatori:2009iz,Hristov:2010ri,Dall'Agata:2010gj}.
We hope to come back to these points in a future publication.

\acknowledgments

This work was partially supported by INFN and MIUR-PRIN contract 2009-KHZKRX.

\end{document}